\documentclass[final,5p,times,twocolumn,authoryear]{elsarticle}
\usepackage{graphicx} 
\usepackage{amsmath,amssymb,bm,mathtools}
\usepackage{physics}
\usepackage{braket}
\usepackage{hyperref}
\hypersetup{hidelinks}
\usepackage{enumitem}
\usepackage{tensor}
\usepackage{mathrsfs}
\begin{document}

\begin{frontmatter}

\title{Neural-Network-Based Variational Method in Nuclear Density Functional Theory: Application to the Extended Thomas–Fermi Model}
\author[first]{Kenta Yoshimura}
\affiliation[first]{organization={Institute of Science Tokyo},
            addressline={yoshimura.k.af21@m.isct.ac.jp}, 
            city={Meguro},
            postcode={152-8550}, 
            state={Tokyo},
            country={Japan}}

\begin{abstract}
We propose a neural-network-based variational framework for nuclear Density Functional Theory based on the extended Thomas--Fermi (ETF) model, in which proton and neutron number densities are represented by multilayer perceptrons and determined by direct minimization of a Skyrme-type energy density functional.
We clarify the mathematical connection to the conventional Euler--Lagrange formulation, showing that stationarity in parameter space corresponds to a projected Euler--Lagrange condition on the neural-network trial-density manifold.
The basic validity of the framework is examined through three sets of calculations: a Woods--Saxon potential benchmark, ground-state calculations of finite nuclei ($^{40}$Ca, $^{90}$Zr, and $^{208}$Pb), and nuclear pasta phases.
The binding energies of finite nuclei agree with existing ETF calculations to within $0.5\%$, and representative pasta structures including spheres, rods, and slabs are reproduced.
We also find that single-precision arithmetic yields results comparable to double precision, suggesting that the present framework is well suited to GPU environments in which low-precision computation is advantageous.
\end{abstract}

\begin{keyword}
Nuclear density functional theory\sep Extended Thomas--Fermi\sep Neural networks\sep Variational methods\sep Nuclear pasta
\end{keyword}

\end{frontmatter}


\section{Introduction}
\label{sec:introduction}

Density functional theory (DFT) in nuclear physics \citep{nakatsukasa2016, Colo2020} provides a powerful framework for the unified description of nuclear many-body phenomena ranging from the structure of finite nuclei to the properties of neutron-star matter \citep{Bender2003, Yang2020}.
The conceptual foundation of DFT is the Hohenberg--Kohn (HK) theorem \citep{HK1964}: (i) the ground-state density uniquely determines all ground-state properties, and (ii) the true ground state is obtained by variationally minimizing the exact energy density functional (EDF).
As the standard practical realization, the Kohn--Sham (KS) scheme \citep{KS1965} introduces auxiliary single-particle orbitals to incorporate kinetic energy and shell effects efficiently.
As nuclear DFT is applied to an ever-broader range of phenomena, from spectroscopic studies of individual nuclei to neutron-star matter and complex three-dimensional nuclear pasta structures, the demand for computationally efficient and scalable implementations has grown accordingly.

In parallel with these nuclear physics developments, high-performance computing (HPC) platforms are undergoing a fundamental architectural shift toward GPU- and AI-accelerator-based architectures, which natively support the tensor operations and mixed-precision arithmetic on which modern machine-learning (ML) frameworks are based \citep{Micikevicius2018, Mittal2019, Micikevicius2022, Kashi2024, Silvano2025}.
This hardware evolution has catalyzed a broad \textit{AI for science} movement, in which scientific computing codes are redesigned to leverage the software and hardware infrastructure of the ML ecosystem \citep{jordan2015, lecun2015}.
For nuclear DFT to fully benefit from these new computational architectures, it is important to develop frameworks that are compatible with GPU-based ML libraries, where gradient computation through automatic differentiation and parameter optimization through backpropagation are fundamental operations.

A common strategy for applying AI to physics is to train a surrogate model that emulates the input--output behavior of an existing simulator or a governing equation.
On the other hand, DFT offers a different opportunity because its basic formulation is already variational.
According to the variational principle underlying the HK theorem, the ground state is obtained merely by minimizing the functional over admissible densities under the appropriate constraints.
This variational structure suggests a natural connection between DFT and ML-based optimization: the EDF can be treated as the objective function, while the density distribution is represented as the model output.
In this formulation, gradient-based optimization can be used to search for the minimizing density within a chosen variational ansatz, without replacing the EDF minimizationby a separately trained surrogate model.
Conventional implementations of nuclear DFT have realized this variational principle \textit{indirectly} by deriving and implementing the corresponding mean-field or Euler--Lagrange equations and then solving them in a self-consistent manner, requiring algorithms tailored to the physical structure of the problem, such as the construction of mean-field Hamiltonians, the solution or diagonalization of single-particle equations, and the iterative adjustment of particle-number constraints through chemical potentials.
The direct variational approach replaces the explicit construction of these equations with automatic differentiation and gradient-based parameter optimization, thereby providing a unified computational route for EDF minimization once the functional and constraints are specified.
Neural networks, in particular multilayer perceptrons (MLPs), provide a natural and flexible parameterization for the trial density, as they can serve as universal function approximators \citep{hornik1989multilayer}.
This flexibility may make it possible to represent density distributions ranging from spherical finite nuclei to complex three-dimensional pasta structures without imposing any symmetry constraint a priori, while standard ML libraries such as PyTorch provide GPU-accelerated automatic differentiation that efficiently computes gradients through backpropagation.
In electronic-structure physics, related synergies between energy minimization and NN-based optimization have already been demonstrated: FermiNet \citep{Pfau2020} and PauliNet \citep{Hermann2020} achieve highly accurate ground-state energies by directly minimizing variational energy functionals with neural-network trial wave functions, and similar ideas underlie neural-network quantum states \citep{Carleo2017} and ML-based functional improvements \citep{Kirkpatrick2021}.
However, the direct use of NNs as a variational tool for EDF minimization in nuclear DFT remains largely unexplored.

As a first step toward realizing this framework, the Extended Thomas--Fermi (ETF) approximation \citep{Bartel1985, dutta1986, centelles1990self, centelles2007thomas} offers a particularly natural starting point.
ETF is one of the most established orbital-free implementations of nuclear EDFs: the EDF is expressed entirely in terms of the proton and neutron number densities and their spatial derivatives via a semiclassical expansion, without introducing any single-particle orbitals.
Although ETF does not treat shell effects exactly, it has demonstrated practical utility across a wide range of systems, including finite nuclei, neutron-star equations of state \citep{Pearson2018, shchechilin2024, chamel2025}, and large-scale three-dimensional nuclear pasta phases \citep{nakamuramaster, nakamura_prep}.
Additionally, there have been proposed further extensions for including shell effects by means of non-local contributions as well as machine-learning method~\cite{wu2022,wu2025,wu2026}.
In ETF, the variational problem therefore reduces to finding two density distributions that minimize the functional under particle-number constraints, which provides a natural setting for representing the densities directly as outputs of neural networks (NNs).
By contrast, extending this approach to KS, HF, or HFB theory would require additional ingredients, such as NN representations of single-particle orbitals or quasiparticle wave functions, orthonormality constraints, and pairing degrees of freedom, making the problem considerably more involved.
ETF thus serves as a natural first testbed for establishing and validating the methodology before extending it to more microscopic levels of nuclear theory.

In this paper, we construct an NN-based variational framework in which the proton and neutron density distributions are represented by MLPs and the Skyrme+ETF energy density functional is minimized directly as the loss function.
We clarify the mathematical connection to the conventional Euler--Lagrange formulation, showing that stationarity in parameter space corresponds to a projected Euler--Lagrange condition on the NN trial-density manifold.
The basic validity of the framework is examined through three sets of calculations: a Woods--Saxon potential benchmark, ground-state calculations of finite nuclei ($^{40}$Ca, $^{90}$Zr, and $^{208}$Pb), and nuclear pasta phases.
We also find that, in the present calculations, single-precision arithmetic yields results comparable to double precision, suggesting that the present framework is well suited to GPU environments in which low-precision computation is advantageous.

The organization of this paper is as follows.
In Sec.~\ref{sec:formulation}, we present the Skyrme+ETF energy density functional and the formulation of the NN-based variational optimization.
In Sec.~\ref{sec:results}, we present and discuss the benchmark, finite-nuclei, and pasta-phase calculations.
Sec.~\ref{sec:summary} is dedicated to show the summary and outlook.


\section{Formulation}
\label{sec:formulation}

\subsection{NN-Based Variational Method for the ETF Model}

In the present ETF calculation, the objective is to determine the ground-state proton and neutron density distributions by minimizing the energy functional with respect to $n_p(\bm{r})$ and $n_n(\bm{r})$ under fixed particle-number constraints.
If the physical energy functional is denoted simply by $E[n_n,n_p]$, the variational problem is formulated as
\begin{equation}
    \min_{n_p,n_n} E[n_n,n_p]
\end{equation}
subject to
\begin{equation}
    \int_\Omega d\bm{r}\, n_p(\bm{r}) = Z,
    \qquad
    \int_\Omega d\bm{r}\, n_n(\bm{r}) = N,
    \label{eq:numberconstraint}
\end{equation}
where $Z$ and $N$ are the proton and neutron numbers, respectively, and $\Omega$ denotes the computational domain.

By introducing Lagrange multipliers $\mu_q$, the corresponding Euler--Lagrange equations are written as
\begin{equation}
    \frac{\delta E}{\delta n_q(\bm{r})} = \mu_q,
    \qquad q \in \{n,p\}.
    \label{eq:EL_original}
\end{equation}
In conventional approaches, these equations are derived analytically for a given functional and then solved self-consistently.
In the present work, by contrast, we do not explicitly construct the Euler--Lagrange equations.
Instead, we represent the density distributions directly by neural networks (NNs) and perform the minimization in parameter space.

For each particle species $q$, we introduce a multilayer perceptron (MLP) $g_q(\bm{r};\theta_q)$ that takes the spatial coordinate $\bm{r}$ as input~\citep{hornik1989multilayer}, and define the nonnegative unnormalized density by
\begin{equation}
    \hat n_q(\bm{r};\theta_q) = \exp\qty(g_q(\bm{r};\theta_q)).
    \label{eq:raw_density}
\end{equation}
With this construction, nonnegativity of the density is guaranteed automatically.
The normalized density is then defined as
\begin{equation}
    n_q(\bm{r};\theta_q)
    =
    N_q
    \frac{\hat n_q(\bm{r};\theta_q)}
    {\int_\Omega d\bm{r}'\, \hat n_q(\bm{r}';\theta_q)}
    \label{eq:normalized_density}
\end{equation}
so that the particle-number constraints in Eq.~\eqref{eq:numberconstraint} are satisfied identically at every optimization step.

Let the whole parameter space be $\Theta=\Theta_n\times\Theta_p$.
The trial-density manifold induced by the NN ansatz is written as
\begin{equation}
    \mathcal{M}_{\Theta}
    =
    \left\{
    \bigl(n_p(\cdot;\theta_p),\, n_n(\cdot;\theta_n)\bigr)
    \;\middle|\;
    (\theta_n,\theta_p)\in\Theta
    \right\}.
    \label{eq:trial_manifold}
\end{equation}
The variational problem is thereby reduced to the optimization problem
\begin{equation}
    \min_{(\theta_n,\theta_p)\in\Theta}
    \mathcal{L}(\theta_n,\theta_p),
    \label{eq:theta_optimization}
\end{equation}
where $\mathcal{L}$ denotes the total loss function.
In the main optimization stage, the loss is set to the total energy of the system, namely $\mathcal{L}=E$.
However, we note that, during the initial stage of the optimization, an auxiliary guiding term may be added to stabilize the density profile, as described in Sec.~\ref{sec:guide}.

Assuming that $E$ is differentiable with respect to the densities, the gradient with respect to a network parameter $\theta_{q,i}$ is given by the chain rule,
\begin{equation}
    \pdv{E}{\theta_{q,i}}
    =
    \int_\Omega d\bm{r}\,
    \frac{\delta E}{\delta n_q(\bm{r})}
    \pdv{n_q(\bm{r};\theta_q)}{\theta_{q,i}}.
\end{equation}
At a stationary point of Eq.~\eqref{eq:theta_optimization}, this gradient vanishes for all $i$:
\begin{equation}
    \int_\Omega d\bm{r}\,
    \frac{\delta E}{\delta n_q(\bm{r})}
    \pdv{n_q(\bm{r};\theta_q)}{\theta_{q,i}}
    = 0
    \label{eq:projected_EL}
\end{equation}
This means that the functional derivative $\delta E/\delta n_q$ is orthogonal to all tangent vectors $\partial n_q/\partial\theta_{q,i}$ of the trial manifold $\mathcal{M}_{\Theta}$.
Since the normalization condition in Eq.~\eqref{eq:normalized_density} implies
\begin{equation}
    \int_\Omega d\bm{r}\,
    \pdv{n_q(\bm{r};\theta_q)}{\theta_{q,i}} = 0,
\end{equation}
one may add an arbitrary constant $\mu_q$ without changing Eq.~\eqref{eq:projected_EL}, and obtain
\begin{equation}
    \int_\Omega d\bm{r}\,
    \left(
    \frac{\delta E}{\delta n_q(\bm{r})}
    - \mu_q
    \right)
    \pdv{n_q(\bm{r};\theta_q)}{\theta_{q,i}}
    = 0
    \label{eq:projected_EL_mu}
\end{equation}
Equation~\eqref{eq:projected_EL_mu} may therefore be interpreted as the Euler--Lagrange condition projected onto the tangent space of the NN trial manifold.
In this sense, the present method constitutes a realization of the Ritz variational principle formulated in a neural-network function space.
When the trial manifold is sufficiently expressive, the solution obtained in this way is expected to approach the stationary solution of the original constrained variational problem.

In the actual numerical implementation, the continuous expressions above are evaluated on a three-dimensional mesh.
That is, the spatial integrals are replaced by discrete sums over the grid points, and the energy is minimized with respect to the NN parameters through backpropagation on the discretized computational domain.

\subsection{Skyrme+ETF Energy Density Functional}

In the present work, the total energy is written as
\begin{equation}
    E[n_n,n_p]
    =
    E_{\mathrm{kin}}^{\mathrm{eff}}
    +
    E_{\mathrm{int}}
    +
    E_{\mathrm{Coul}},
    \label{eq:Etotal}
\end{equation}
where $E_{\mathrm{kin}}^{\mathrm{eff}}$ is the kinetic energy after the center-of-mass correction is subtracted,
$E_{\mathrm{int}}$ denotes the contribution from the Skyrme interaction,
and $E_{\mathrm{Coul}}$ is the Coulomb energy.
The explicit expressions of these terms are given below.

\subsubsection{Kinetic Energy}

The kinetic energy is given by
\begin{equation}
    E_{\mathrm{kin}}
    =
    \sum_{q=n,p}
    \frac{\hbar^2}{2m_q}
    \int_\Omega d\bm{r}\, \tau_q(\bm{r}),
    \label{eq:Ekin}
\end{equation}
where $\tau_q(\bm{r})$ is the kinetic-energy density.
Within the ETF approximation, it is decomposed as
\begin{equation}
    \tau_q
    =
    \tau_q^{\mathrm{TF}}
    +
    \tau_q^{(2)}
    +
    \tau_q^{(4)},
    \label{eq:tau_decomp}
\end{equation}
with the Thomas--Fermi term
\begin{equation}
    \tau_q^{\mathrm{TF}}
    =
    \frac{3}{5}(3\pi^2)^{2/3} n_q^{5/3}.
    \label{eq:tauTF}
\end{equation}

The second-order ETF correction is written as
\begin{equation}
\begin{aligned}
    \tau_q^{(2)}(\bm{r})
    &=
    \frac{1}{3}\Delta n_q
    +\frac{1}{36}\frac{|\nabla n_q|^2}{n_q}
    +\frac{1}{6}\frac{\nabla f_q \cdot \nabla n_q}{f_q}
    +\frac{1}{6}n_q\frac{\Delta f_q}{f_q} \\
    &\quad
    -\frac{1}{12}n_q\qty(\frac{\nabla f_q}{f_q})^2
    +\frac{1}{2}\qty(\frac{2m_q}{\hbar^2})^2
    n_q
    \qty(\frac{|\bm{W}_q|}{f_q})^2,
\end{aligned}
\label{eq:tau2}
\end{equation}
where
\begin{equation}
    f_q(\bm{r})
    \equiv
    \frac{m_q}{m_q^*(\bm{r})}
    =
    1+\frac{2m_q}{\hbar^2}\qty(B_3 n(\bm{r}) + B_4 n_q(\bm{r}))
    \label{eq:fq}
\end{equation}
is the effective-mass ratio which has the isoscalar density $n=n_n+n_p$ and the spin-orbit field $\bm{W}_q$ defined later.

For the fourth-order correction, we adopt, for simplicity, the expression derived for the case $f_q=1$:
\begin{equation}
    \tau_q^{(4)}
    =
    (3\pi^2)^{-2/3} n_q^{1/3}
    \qty[
    \frac{1}{270}u_q^2
    -\frac{1}{240}u_q v_q
    +\frac{1}{810}v_q^2
    ],
    \label{eq:tau4}
\end{equation}
with
\begin{equation}
    u_q = \frac{\Delta n_q}{n_q},
    \qquad
    v_q = \frac{|\nabla n_q|^2}{n_q^2}.
    \label{eq:uv_def}
\end{equation}
This treatment is the same as in the previous study~\citep{centelles1990self}, which allows direct comparison of the calculation results.

Since $E_{\mathrm{kin}}$ calculated above contains the contribution associated with the center-of-mass motion, we subtract it approximately as
\begin{equation}
    E_{\mathrm{cm}} \approx \frac{E_{\mathrm{kin}}}{A},
    \qquad
    A = \int_\Omega d\bm{r}\, (n_n+n_p),
    \label{eq:Ecm}
\end{equation}
and define the corrected kinetic energy as
\begin{equation}
    E_{\mathrm{kin}}^{\mathrm{eff}}
    =
    E_{\mathrm{kin}} - E_{\mathrm{cm}}.
    \label{eq:Ekin_eff}
\end{equation}
This correction is employed here as a simple approximate treatment of the center-of-mass motion.

\subsubsection{Interaction Energy}

The interaction energy is given by
\begin{equation}
    E_{\mathrm{int}}
    =
    \int_\Omega d\bm{r}\,
    \mathcal{E}_{\mathrm{int}}(\bm{r}),
    \label{eq:Eint}
\end{equation}
where the interaction-energy density is written as
\begin{equation}
\begin{aligned}
\mathcal{E}_{\mathrm{int}}(\bm{r})
&=
\sum_{q=n,p}
\bm{W}_q\cdot\bm{J}_q
+ B_1 n^2
+ B_2 \sum_{q=n,p} n_q^2
+ B_5 n\,\Delta n\\
&\quad
+ B_6 \sum_{q=n,p} n_q\,\Delta n_q
+ B_7 n^{\alpha+2}
+ B_8 n^\alpha \sum_{q=n,p} n_q^2.
\end{aligned}
\label{eq:int_density}
\end{equation}

The spin-orbit field and the spin-current density are defined by
\begin{align}
    \bm{W}_q &= B_9\qty(\nabla n + \nabla n_q),
    \label{eq:Wq}
    \\
    \bm{J}_q &=-\qty(\frac{2m_q}{\hbar^2})
    \frac{n_q\,\bm{W}_q}{f_q}.
    \label{eq:Jq}
\end{align}
Accordingly, the functional adopted here is expressed solely in terms of the densities and their spatial derivatives.

\subsubsection{Coulomb Energy}

The Coulomb energy is decomposed into the direct and exchange contributions as
\begin{equation}
    E_{\mathrm{Coul}}
    =
    E_{\mathrm{Coul}}^{\mathrm{dir}}
    +
    E_{\mathrm{Coul}}^{\mathrm{ex}}.
    \label{eq:ECoul}
\end{equation}

The direct term is written in terms of the charge density
\begin{equation}
    n_{\mathrm{ch}}(\bm{r}) = n_p(\bm{r}) - n_e(\bm{r})
    \label{eq:nch}
\end{equation}
as
\begin{equation}
    E_{\mathrm{Coul}}^{\mathrm{dir}}
    =
    \frac{1}{2}
    \int_\Omega d\bm{r}\,
    n_{\mathrm{ch}}(\bm{r})V_{\mathrm{Coul}}(\bm{r}),
    \label{eq:ECoul_dir}
\end{equation}
where $V_{\mathrm{Coul}}(\bm{r})$ satisfies the Poisson equation
\begin{equation}
    \Delta V_{\mathrm{Coul}}(\bm{r})
    =
    -4\pi e^2 n_{\mathrm{ch}}(\bm{r}).
    \label{eq:Poisson}
\end{equation}

For finite nuclei, one simply sets $n_e=0$.
By contrast, when a uniform electron background is assumed, as in neutron-star crust matter, $n_e$ is fixed by the condition of charge neutrality.
The Poisson equation is solved by means of the fast Fourier transform (FFT) algorithm.
For finite nuclei, however, the boundary condition appropriate for an isolated system must be reproduced.
To this end, a procedure that suppresses the effects of periodic images is employed, such as a truncated Coulomb kernel or an equivalent method~\citep{ismailbeigi2006truncation,jin2021a}.

The exchange term is treated within the Slater approximation:
\begin{equation}
    E_{\mathrm{Coul}}^{\mathrm{ex}}
    =
    -\frac{3}{4}\qty(\frac{3}{\pi})^{1/3}
    e^2
    \int_\Omega d\bm{r}\, n_p(\bm{r})^{4/3}.
    \label{eq:ECoul_ex}
\end{equation}

For neutron-star matter calculations, the contribution from the uniform electron gas also appears.
However, when the average proton fraction is fixed, the electron contribution does not affect the variational optimization with respect to the nucleon densities except for an additive constant.
For this reason, the electron energy is omitted from the present functional.

\subsection{Guide Potential and Optimization Strategy}
\label{sec:guide}

In the actual optimization, an auxiliary guiding term is introduced in addition to the physical energy $E$ in order to suppress an excessive spreading of the density distribution during the early stage of training.
The total loss is then defined as
\begin{equation}
    \mathcal{L}
    =
    E
    + w_{\mathrm{guide}} E_{\mathrm{guide}},
    \label{eq:loss_total}
\end{equation}
where
\begin{equation}
    E_{\mathrm{guide}}
    =
    \int_\Omega d\bm{r}\,
    n(\bm{r})V_{\mathrm{guide}}(\bm{r}),
    \qquad
    n=n_n+n_p,
    \label{eq:Eguide}
\end{equation}
and $w_{\mathrm{guide}}\ge 0$ is the weight assigned to the guiding term.

The specific form of $V_{\mathrm{guide}}$ depends on the system under consideration.
For finite nuclei, an attractive Woods--Saxon-type potential may be used, whereas for pasta phases a guiding potential adapted to the target geometry is introduced.
Typical examples of such guiding potentials are summarized in \citep{schuetrumpf2015}.
We emphasize, however, that the final energies and density distributions reported in this work are evaluated only after sufficient convergence has been reached under the condition
\begin{equation}
    w_{\mathrm{guide}}=0.
\end{equation}
The guiding term therefore serves only as an auxiliary device for stabilizing the early stage of the optimization and is not included in the physical energy itself.

\subsection{Computational Settings}

\begin{table*}[t]
\centering
\caption{Benchmark results for the WS potential. 
The total energy (in MeV) is averaged over the last $1000$ steps of a $100000$-step optimization. 
The computational time (in seconds) corresponds to $1000$ optimization steps.}
\label{tab:benchmark_ws}
\begin{tabular*}{0.9\linewidth}{@{\extracolsep{\fill}}c|ccc||c|ccc}
\hline
\multicolumn{4}{c||}{Single precision} & 
\multicolumn{4}{c}{Double precision} \\
\hline
$N_{\rm perc}$ & Energy & H100 & RTX 5000 Ada &
$N_{\rm perc}$ & Energy & H100 & RTX 5000 Ada \\
\hline
16  & -1320.71   & 4.340 & 3.127 & 
16  & -1320.8501 & 4.064 & 4.079 \\
32  & -1321.37   & 4.224 & 3.030 & 
32  & -1321.18   & 4.346 & 5.139 \\
64  & -1321.78   & 4.095 & 3.441 & 
64  & -1321.87   & 4.126 & 8.699 \\
128 & -1322.20   & 4.053 & 3.865 & 
128 & -1322.06   & 4.171 & 20.36 \\
256 & -1321.82   & 4.689 & 5.710 & 
256 & -1322.04   & 5.662 & 65.73 \\
\hline
\end{tabular*}
\end{table*}

In the present work, the density distributions on a three-dimensional mesh are represented by neural networks.
A typical numerical setup is
\begin{equation}
    dx=dy=dz=0.5~\mathrm{fm},
    \qquad
    N_x=N_y=N_z=64,
\end{equation}
which corresponds to a computational box of size
\begin{equation}
    L_x=L_y=L_z=32~\mathrm{fm}.
\end{equation}

For each particle species, the NN is taken to be an MLP with input $\bm{r}=(x,y,z)$, and a representative architecture is
\begin{equation}
    3 \to N_{\mathrm{perc}} \to N_{\mathrm{perc}} \to 1,
\end{equation}
where $N_{\mathrm{perc}}=64$ is used in typical calculations.
The activation function in the hidden layers is chosen to be $\tanh$.
With the construction in Eqs.~\eqref{eq:raw_density} and \eqref{eq:normalized_density}, both nonnegativity of the density and exact particle-number conservation are automatically satisfied.

During training, the loss in Eq.~\eqref{eq:loss_total} is evaluated over all grid points simultaneously, and the NN parameters are updated by backpropagation.
The present method therefore corresponds to a full-batch optimization on the entire mesh.
The Adam algorithm is employed as the optimizer with learning rate
\begin{equation}
    \eta = 10^{-3}.
\end{equation}
The total number of optimization steps is typically taken to be
\begin{equation}
    N_{\mathrm{step}} = 50000
\end{equation}
for finite nuclei and
\begin{equation}
    N_{\mathrm{step}} = 100000
\end{equation}
for pasta phases.

In addition, a pretraining stage may be introduced in order to initialize the NN with a localized density profile.
At this stage, a Woods--Saxon-type distribution~\citep{woods1954diffuse},
\begin{equation}
    n_q^{\mathrm{WS}}(\bm{r})
    =
    \frac{n_{0,q}}
    {1+\exp\qty(\frac{r-R_q}{a_q})},
    \label{eq:WS_density}
\end{equation}
is employed as the target density, and the pretraining loss is defined as
\begin{equation}
    \mathcal{L}_{\mathrm{pre}}
    =
    \sum_{q=n,p}
    \int_\Omega d\bm{r}\,
    \qty(
    n_q(\bm{r};\theta_q)-n_q^{\mathrm{WS}}(\bm{r})
    )^2.
    \label{eq:pretrain_loss}
\end{equation}
Here, $R_q$ and $a_q$ are chosen according to the target nucleus, and $n_{0,q}$ is determined so as to satisfy the normalization condition.
Thus, pretraining serves to provide a physically reasonable initial density profile, whereas the guiding term in Eq.~\eqref{eq:loss_total} is used to stabilize the early stage of the subsequent energy minimization.

The weight of the guiding term, $w_{\mathrm{guide}}$, is initially set to unity and then gradually reduced during the first $N_{\mathrm{ann}}$ optimization steps.
A representative choice is
\begin{equation}
    w_{\mathrm{guide}}(s)
    =
    \max\qty(1-\frac{s}{N_{\mathrm{ann}}},\,0),
    \qquad
    N_{\mathrm{ann}}=10000,
    \label{eq:wguide_schedule}
\end{equation}
so that $w_{\mathrm{guide}}=0$ for all $s>N_{\mathrm{ann}}$.
Beyond that point, the optimization proceeds solely through the minimization of the physical energy $E$.

\section{Results and Discussion}
\label{sec:results}

\begin{table*}[t]
\centering
\caption{Comparison of the present calculation with the SkM$^{*}$ TF$\hbar^4$ results of Ref.~\citep{centelles1990self}.}
\label{tab:compare_centelles_table9}
\begin{tabular*}{0.7\linewidth}{@{\extracolsep{\fill}}c|ccc|ccc}
\hline\hline
Nucl
& \multicolumn{3}{c|}{Present}
& \multicolumn{3}{c|}{Ref.} \\
& $BE$ (MeV) & $r_n$ (fm) & $r_p$ (fm)
& $BE$ (MeV) & $r_n$ (fm) & $r_p$ (fm) \\
\hline
$^{40}$Ca  & $-348.87$ & 3.307 & 3.348 & $-349.8$ & 3.33 & 3.37 \\
$^{90}$Zr  & $-790.73$ & 4.263 & 4.196 & $-792.2$ & 4.27 & 4.20 \\
$^{208}$Pb & $-1630.8$ & 5.614 & 5.467 & $-1631.9$ & 5.62 & 5.47 \\
\hline\hline
\end{tabular*}
\end{table*}

In this section, we examine the validity and applicability of the proposed method from several viewpoints.
First, in order to confirm to what extent the present framework can reproduce existing ETF-based calculations, we calculate the ground state in a Woods--Saxon potential and compare the obtained results with those of previous studies.
Next, for the finite nuclei $^{40}$Ca, $^{90}$Zr, and $^{208}$Pb, we compare the binding energies and the proton and neutron radii with those obtained in previous ETF calculations~\citep{centelles1990self}.
Finally, we present the calculated results for pasta phases and investigate whether various pasta structures such as spheres, rods, and slabs can also be properly described within the neural-network-based density representation adopted in this work.

\subsection{Benchmark}
In this subsection, we discuss the calculation results for a simple Woods--Saxon potential.
In addition, calculations are performed on multiple GPU architectures to investigate how the computational time and the results depend on the number of perceptrons $N_\text{perc}$, as well as how they are affected when the numerical precision of the variables is reduced to single precision.

Table \ref{tab:benchmark_ws} shows the benchmark results obtained on two GPU architectures, NVIDIA H100 and RTX 5000 Ada.
For several values of the perceptron number $N_\text{perc}$, the optimization was carried out for $10^5$ steps; the total energy shown in the table is the average over the last $10^3$ steps, and the computational time is given per $10^3$ optimization steps.
The left half of the table corresponds to calculations in which all variables are defined in single precision, whereas the right half corresponds to calculations in double precision.
The reference value from the previous study used for comparison is $-1323.2$ MeV \citep{centelles1990self}.

At least three points can be identified from this table.
In the first place, although the error is somewhat larger for $N_\text{perc}=16$, the total energy approaches the reference value as $N_\text{perc}$ increases, and overall the deviation remains at the level of about $10^{-1}\%$ relative to the reference value.
Therefore, at least for this benchmark problem, the three-dimensional density representation based on a neural network provides a ground state consistent with that obtained in conventional ETF calculations.
In the second place, comparison between the single- and double-precision results shows that the difference in the obtained total energies is small.
Thus, within the scope of this benchmark, sufficiently good reproducibility is achieved even when single precision is used.
This suggests that the accuracy of the result depends more strongly on the freedom available to represent the density distribution, namely the size of $N_\text{perc}$, than on the difference in numerical precision itself.
This finding is consistent with the compatibility of the present framework with GPU and AI-accelerator hardware, where single-precision arithmetic is natively supported and generally more efficient than double precision.
In the third place, a clear difference is observed in the computational time between the H100 and the RTX 5000 Ada.
For the RTX 5000 Ada, the performance was comparable to, or in some cases even better than, that of the H100 when $N_{\text{perc}}$ was small.
However, in double precision or for larger $N_{\text{perc}}$, the computational time increased substantially, particularly for $N_{\text{perc}}=128$ and 256.
By contrast, for the H100, the computational time exhibited only limited variation with changes in precision and $N_{\text{perc}}$.
These results indicate that the dependence of computational cost on numerical precision and network size differs between the two GPUs.
Therefore, in the present method, it is important to select an implementation and computational environment suited to the GPU in use in order to achieve both high expressive power and computational efficiency.

\subsection{Finite Nuclei}
In this subsection, we compare the calculated results for finite nuclei with the TF$\hbar^4$ results for SkM$^{*}$ reported in a previous study \citep{centelles1990self}.
Table \ref{tab:compare_centelles_table9} shows the comparison of the binding energies and the neutron and proton radii for $^{40}$Ca, $^{90}$Zr, and $^{208}$Pb.

These results show that the NN-based calculations are in good agreement with the previous ETF calculations.
More specifically, the difference in the binding energy is at most about $0.5\%$, while the radii differ by approximately $1\%$ or less.
Therefore, the present NN-based framework can describe the density distributions of finite nuclei, as well as the basic physical quantities derived from them, with sufficient accuracy.

One possible origin of the remaining differences is the difference in the discretization conditions used in the calculations.
In the previous study, a one-dimensional calculation assuming spherical symmetry was performed with a mesh spacing of $\Delta r = 0.1\,\mathrm{fm}$.
In contrast, the present study employs a three-dimensional mesh without imposing any symmetry, with a grid spacing of $\Delta x = 0.5\,\mathrm{fm}$.
Such differences in the numerical setup may affect the density distribution, especially in the surface region, and consequently the evaluated radii.
Therefore, the differences seen in Table \ref{tab:compare_centelles_table9} can be regarded as acceptable when the differences in dimensionality and discretization conditions are taken into account.

\subsection{Pasta Phases}
In this subsection, we present the calculation results for nuclear pasta phases.
At the initial stage of the learning procedure, an external potential $V_{\mathrm{guide}}(\bm{r})$ corresponding to the target geometry is introduced in order to provide an initial bias toward representative structures such as spheres, rods, and slabs.
For the explicit form of the guide potential, see Ref.~\citep{schuetrumpf2015, schuetrumpf2019}.
In standard calculations of pasta phases, it is necessary to perform calculations for various cell sizes and to find the cell size at which the energy is minimized in order to determine the most stable structure.
However, the purpose of this subsection is simply to examine whether the present NN-based framework can reproduce pasta phases.
Therefore, the calculations in this subsection are carried out with the cell size fixed at $L_x=L_y=L_z=32\,\mathrm{fm}$, as in the previous calculations.

\begin{figure}
    \centering
    \includegraphics[width=0.9\linewidth]{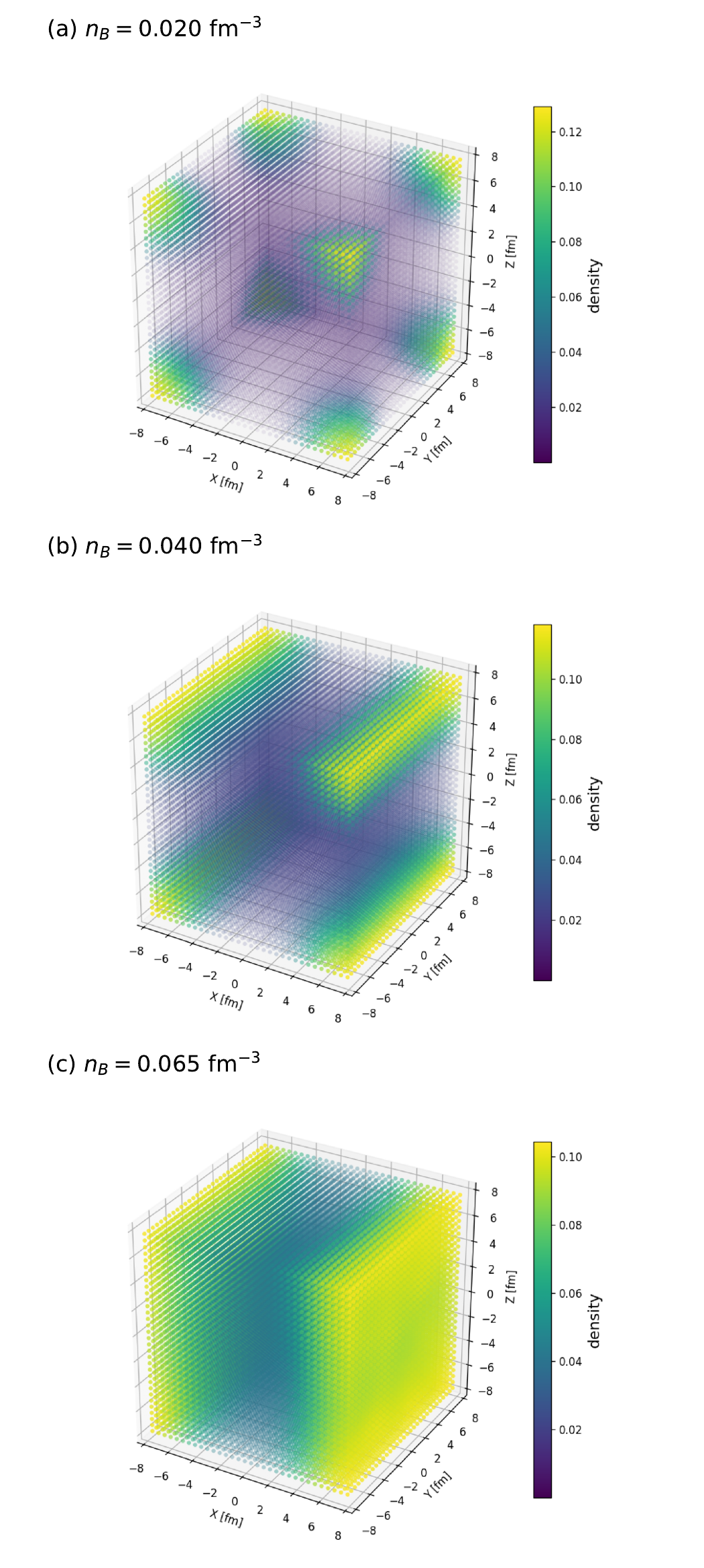}
    \caption{Density distributions obtained for representative pasta phases at a fixed proton fraction $Y_p=0.1$. The figure shows that the present neural-network-based calculation reproduces the characteristic sphere, rod, and slab structures for different values of $n_B$.}
    \label{fig:pasta_Yp01}
\end{figure}

Figure \ref{fig:pasta_Yp01} shows the density distributions obtained at a fixed proton fraction for $n_B=0.02,\,0.04,\,0.065\,\mathrm{fm}^{-3}$, where calculations were performed with different guided shapes.
As can be seen from the figure, under each condition the NN-based density representation reproduces the basic structures of nuclear pasta phases, namely spheres, rods, and slabs.
This result indicates that the present method is applicable not only to localized structures such as finite nuclei, but also to more complicated non-spherical density distributions with different topologies.

These results confirm that the flexible, symmetry-unconstrained NN density representation is effective for describing not only spherical finite nuclei but also the topologically diverse structures encountered in nuclear pasta phases.

\section{Summary and Outlook}
\label{sec:summary}

In this work, we have proposed a neural-network-based variational framework for nuclear DFT calculations and demonstrated its basic validity in the context of the ETF model.
In the present framework, the Skyrme-type ETF energy density functional serves directly as the loss function, and the proton and neutron density distributions are parameterized by MLPs and determined by gradient-based optimization, without analytically deriving or solving the corresponding Euler--Lagrange equations.
We have clarified the mathematical connection to the conventional variational formulation: stationarity in the NN parameter space corresponds to a projected Euler--Lagrange condition on the trial-density manifold and can be interpreted as a Ritz variational principle in neural-network function space.
This formulation provides a practical way to exploit the variational structure underlying DFT within a modern ML-based optimization framework.

The framework has been examined through three sets of calculations.
In the benchmark calculation using a Woods--Saxon potential, the total energy agrees with the reference ETF value to within $10^{-1}\%$, demonstrating the basic accuracy of the NN-based variational optimization.
For the finite nuclei $^{40}$Ca, $^{90}$Zr, and $^{208}$Pb, the binding energies agree with existing ETF calculations within $0.5\%$, and the proton and neutron radii are reproduced within approximately $1\%$.
The calculations of nuclear pasta phases further demonstrate that the NN density representation, without any symmetry constraint imposed a priori, can describe a wide range of three-dimensional density structures, from spherical nuclei to rods and slabs.

A finding of direct relevance to GPU/AI-accelerator-oriented computing environments is that, in the present calculations, reducing the numerical precision to single precision yields results comparable to those obtained in double precision.
This suggests that the present framework is well suited to GPU and AI-accelerator hardware, where low-precision arithmetic often provides substantial computational advantages, and that nuclear DFT calculations can be carried out within the ML ecosystem that drives current AI for science developments.

A practical advantage of the present approach is that, once a functional is specified, the corresponding Euler--Lagrange equations need not be derived analytically on a case-by-case basis.
In conventional approaches, this derivation becomes increasingly cumbersome as the functional is extended to include higher-order gradient corrections or finite-temperature effects \citep{Brack1985, centelles1990self, centelles2007thomas}.
The present framework, based on automatic differentiation, provides a systematic way to incorporate such extensions within the same computational scheme once the corresponding functional is implemented, making it natural to explore higher-order ETF corrections or finite-temperature ETF \citep{brack1984} as immediate future directions.
A further important direction is to extend the present approach beyond the ETF level to more microscopic mean-field theories, such as Skyrme Hartree--Fock and Hartree--Fock--Bogoliubov.
Such an extension would require representing single-particle orbitals, quasiparticle wave functions, or density matrices by NN ansatzes, together with orthonormality and pairing constraints.
Although electronic-structure methods such as FermiNet \citep{Pfau2020} and PauliNet \citep{Hermann2020} are not direct analogues of nuclear EDF solvers, they provide important precedents for NN-based variational representations of quantum many-body states.
Nuclear systems pose additional challenges, including spin-orbit couplings, density-dependent terms, pairing correlations, and possible extensions involving three-body forces, but realizing this direction would open a new computational path for nuclear many-body theory, bridging nuclear physics and computational science.

\section*{Acknowledgement}
The author thanks Yo Nakamura for invaluable discussions and information of the three-dimensional ETF calculations.
This author is financially supported by the JSPS Research Fellow, Grant No.~JP24KJ1110.
This work used the computational resources of TSUBAME4.0 at Institute of Science Tokyo (Project ID: hp240183, hp250097), and Miyabi at Joint Center for Advanced High Performance Computing (JCAHPC) in the University of Tokyo (project ID: hp260222), through the HPCI System Project.

\bibliographystyle{elsarticle-harv} 
\bibliography{etfml}

\end{document}